# Nanoscale ear drum: Graphene based nanoscale sensors

*S. M. Avdoshenko\*, C. G. Rocha\*, G. Cuniberti,*

[*]     Dr. S. M. Avdoshenko , Dr. C. G. Rocha
Address  Institute for Materials Scienceand Max Bergmann Center of Biomaterials
Address  Dresden University of Technology , D-01062
E-mail: savdoshenko@nano.tu-dresden.de
E-mail: cgroacha@nano.tu-dresden.de
        Prof. Dr. G.Cuniberti
Address Institute for Materials Scienceand Max Bergmann Center of Biomaterials
Address Dresden University of Technology , D-01062
E-mail: gcuneberti@nano.tu-dresden.de

CGR would like to thank Alexander von Humboldt Foundation. SMA is thankful for financial support from the Erasmus Mundus programme External Co-operation (EM ECW-L04 TUD 08-11)

The difficulty in determining the mass of a sample increases as its size diminishes. At the nanoscale, there are no direct methods for resolving the mass of single molecules or nanoparticles and so more sophisticated approaches based on electromechanical phenomena are required. More importantly, one demands that such nanoelectromechanical techniques could provide not only information about the mass of the target molecules but also about their geometrical properties. In this case, the performance of such nanoscale sensors is highly dependent on which material is used as host layer. We can say that a promising detector should at least hold simultaneously optimal mechanical and electrical properties. An ideal candidate for such purposes is the one-atom-thick carbon layer known as graphene, the newest member of the "carbon allotrope family". Hailed as the thinnest and strongest material ever made, graphene has revolutionized the scientific frontiers in nanoscience and condensed matter Physics since its first isolation in 2004-2005.[1,2] Since then, a rich variety of remarkable physical features related to graphene structures has been reported, such as high carrier mobility,[3] almost perfect crystalline structure,[4] and anomalous quantum hall effect.[5]

Graphene has overcome the status of simply being an academic model for describing the properties of various carbon-based materials such as graphite, fullerenes and carbon nanotubes. The peculiar electronic and transport characteristics of graphene have guided researchers to explore several important physical phenomena where graphene appears as the main character in investigations focusing on spintronics,[6,7] ac/dc transport,[8,9] and thermoelectrics.[10,11] Additionally, graphene also collects fascinating mechanical properties such as lightness, flexibility and robustness against breakage.[12] Such mechanical and electrical functionalities qualify graphene as an ideal candidate for fabricating nanoelectromechanical systems (NEMS).[13,14]

Recent experimental and theoretical studies have confirmed that the electronic response of graphene samples are rather sensitive to mechanical deformations.[15,16,17] Measurements coupling mechanical and electronic degrees of freedom in graphene include, for instance, detailed analysis of band gap manipulation in uniaxially strained graphene membranes.[18,19] Pushing beyond the scope of electrical sensitivity to force, the holy grail of NEMS applications has been the implementation of nanoscale resonators capable of resolving the mass of single molecules or even single atoms. A new class of NEMS mass sensors was prompted with the design of carbon nanotube resonators anchored to an electrode surface. [20,21,22] Radio frequency waves are applied on the tube, inducing it to vibrate at a certain characteristic resonance frequency. The device can be exposed to other atoms or molecules that attach onto the tube and their presence as well as their mass can be detected through shifts on the resonant frequency. The same route was recently followed using monolayer graphene as resonators operating with electrical readout and at room temperature.[23] Further understanding of the basic features of these devices, including their response with respect to applied voltage, the mass of foreign objects and temperature variations are required in order to boost graphene in nanoscale mass sensor applications. In this letter, we illustrate how graphene membranes can serve as highly sensitive devices capable of measuring the mass and other properties of molecules adsorbed on their surface. Wide graphene sheets were exposed to different types and amounts of molecules and molecular dynamic simulations were employed to treat these doping processes statistically. We demonstrate that the mass variation effect and information about the graphene-molecule interactions can be inferred through dynamical response functions. Our results confirm the potential use of graphene as detector devices with remarkable precision in estimating variations in mass at molecular scale and other physical features of the dopants.

Initially, we introduce our system of study which is a graphene host membrane coupled to electrodes exposed to other molecules. Our results are interpreted based on the same physical principles applied to our ears; for example, one can estimate how large or far away a bouncing ball is only through the sound emitted when it hits the ground. Transferring this mechanism to our graphene based drum, we can say that graphene mimics the eardrum and the electrodes work as our neurons, transmitting the vibrations to the analyzer (brain). We demonstrate that it is possible to build a precise knowledge about the mass and other physical properties of the dopants by analyzing the signals coming from the systematic molecular collisions occurring between the molecules and the graphene membrane.

We performed molecular dynamics (MD) simulations using the classical module FIRST in the CP2k[29] code, which allowed us to obtain relevant physical information in the time domain together with statistical quantities based on the ergodic hypotheses. Each system in our study was explored with 0.5 fs time step and 20 ps total time under the classical force field approximation[25] parametrization created for aromatic carbon (sp2 hybridization) and hydrogen atoms. All aromatic atoms are assumed to be neutral particles. The total time scale was chosen so as to be long enough to guarantee that quick atomic rearrangements were seen. Our mass sensor device is idealized by a fully periodic graphene membrane vibrating at room temperature (T=300 K) which is exposed to foreigner organic molecules. The dimensions of the graphene flake are 60×60 Å$^2$ and as it will be discussed further, the size of the detector dictates how precise is the output of the sensor. Three molecules with different affinities and topologies were chosen to interact with this membrane: (i) coronene (flat macromolecule built with 6 attached benzene rings), (ii) biphenyl (two benzene rings tilted by 30º degrees) and (iii) fullerene (hollow sphere molecule composed purely of carbon atoms). For the sake of simplicity, no explicit environments such as air or solvents were considered.

The dynamic on-fly properties are described by a classical Lagrangian with isothermal constraint conditions given by a Nose-Hoover thermostat[30] applied to a system containing N particles. In non-Hamiltonian formulations, an additional set of phase space coordinates is introduced in the description associated with an effective mass which determines the coupling between the reservoir and the system of interest. The Nose-Hoover thermostat coupling time used in our calculations is 100 fs. In **Figure 1**, we show an example the dynamic trajectories drawn by the molecules deposited on the graphene membrane at room temperature. The shadows capture a sequence of snapshots developed by the three systems of interest: graphene flake interacting with four (a) coronene, (b) biphenyl and (c) fullerene molecules, respectively. Each panel shows a representative trajectory of about 25 ps with 10 fs time sampling between frames. At first glance, one can clearly notice that each molecule develops a different gliding route on the graphene plate. Coronene and biphenyl are rather mobile molecules leaving long tracking paths along the membrane. Buckyballs display slight dislocations on the graphene drum, bouncing mostly in the center of the graphene. We can already infer that the geometrical shapes of the molecules will differently affect the vibrational movements of the graphene drum, offering an effective way to distinguish their physical features and mass.

To underline the real nature of the dynamical processes in our systems, we calculated the time-dependent velocity autocorrelation functions (VAF), C(t), which provides a mean value taken over fluctuation regressions of all occurring initial velocities. Aspects related to the dynamical events can be observed in Fourier transforms of the VAF which can unveil the intrinsically vibrational spectra of the molecular system. Additional information can be obtained by computing the classical dipole autocorrelation functions (CDAF) on its spectral form. This time-dependent autocorrelation function is defined as the Fourier transformation of the dipole moment autocorrelation function, $|d(t)|^2$, which is conceptually similar to Infra-red (IR) spectra limits. The dynamical trajectories are then treated within principal component analysis (PCA) schemes, simplifying considerably the investigation of such complex physical phenomena.

**Figure 2(c)** depicts the Fourier transformation of the VAF obtained firstly for our reference system, which is the pristine graphene membrane when no molecules are included. We split the spectral functions into two main contributions shown by black and red curves, respectively, representing the positive and negative displacements of the atoms with respect to the graphene plane. We also calculated its associated CDAF spectrum which mimics the absorption lineshape of fluctuating charges on the graphene layer under the influence of an incident electric field. We estimate high absorption intensities at the frequency ~ 2000 cm$^{-1}$ which lies within the infra-red range. This frequency is about ~ 400 cm$^{-1}$ far from the experimental value obtained for graphite structures, but still it gives the correct expectation value for the maximum frequency window observed in VAF spectrum. Additionally, the range of the VAF spectrum where most of its features appears (~ 1600 cm$^{-1}$) is in accordance with experimental measurements.

Also, from the velocity spectral functions, one can identify on the inset a pronounced low frequency peak at the frequency of ν ~ 35 cm$^{-1}$. This strong vibrational band corresponds to the fundamental frequency mode of the pristine graphene. By projecting the MD trajectory on the two most significant principle vectors ($V^i$ and $V^{ii}$), one can distinguish these main oscillatory modes. The period of the first fundamental mode ($V^i$) is given by ~ 6 ps and its conversion to frequency corresponds exactly to the value of $v \sim 35$ cm$^{-1}$, which is also observed in VAF spectra function. The second mode was also calculated characterizing the second highest transversal displacements of the graphene membrane with respect to its plane. Such real space characteristics of atomic translocations regarding the first two principal component (PC) vectors are well illustrated on panel **(a)** of the same figure. Regular oscillation patterns can be visualized indicating that the graphene layer apparently behaves as a continuous membrane. It is also straightforward to think that such PC modes have a relation with the membrane tension. In this sense, we expect that these two principal components will be significantly disturbed as molecules are placed on the membrane surface. Although in principle a continuum description could work as a first approximation, we will proceed with our investigation by probing dynamically the sensitivity of the device through a complete atomistic representation.

After analyzing in details the dynamic response of a pristine graphene mat, we are able to identify how these modes are affected by the inclusion of other molecules. As a first example, we consider an isotopic condition where we maintain the same type of molecule (coronene) while changing the amount that is adsorbed on the graphene. **Figure 3 (upper panels)** reveals the trajectory projections on the two first principle component vectors for up to four coronene molecules doping the graphene layer. Red, green and blue curves correspond to situations where one, two, and four molecules stand on the graphene sheet, respectively. One can clearly see that the curves undergo shifts as more molecules interact with graphene indicating that the natural vibrations of the membrane are disturbed in response to modifications in the environment. For a better visualization of such shifts, we converted the time-dependent projections into current spectral functions assuming that our doped graphene membrane behaves as a single-level transmission system. Based on the regular oscillatory patterns seen on **Figure 2(a)**, we represent the graphene detector as a RLC circuit operating at a characteristic resonance frequency of $\omega_o$. We adapted the differential equation of a driven harmonic circuit into our problem where a time dependent power source, $V(t) = E\cos(\omega_{ext}t)$, pumping charges with a frequency of $\omega_{ext}$ will scan the resonance frequency of the pristine and doped structures at ballistic regime. The solution for a RLC system written in terms of frequency is given by

$$I(\omega_0) \sim \frac{E}{\omega_0 \alpha} \times \frac{1}{\sqrt{p^2+1}} \tag{1}$$

where E is the oscillation amplitude of the driven field, $\alpha$ is the attenuation parameter, and $p = (\omega_o - \omega_{ext})/\alpha$.

The estimated current spectral functions are shown on the lower panels of **Figure 3** using the resonance frequencies taken from the PC plots and attenuation of $\alpha = 1.0$ THz. The same color legend used on the upper panels of **Figure 3** is adopted to represent the situations where up to four coronene molecules dope the graphene drum. The dashed lines show the sum over the two components when four molecules stick on the graphene sheet. Looking initially to the current profiles for each component representing the pure graphene membrane, resonance peaks at ~ 1.17 THz and ~ 2.10 THz are found. As more coronene molecules are placed on the membrane, the peaks gradually move upwards in frequency.

In standard MEMS devices the mass loading usually does not affect the mechanical tension of the sensor and so one can say the fundamental frequency of the oscillator changes in terms of mass as $\delta\omega_o = -\delta m_{eff}\omega_o/2m$ being $m_{eff}$ and m the effective mass of the whole doped and pristine membrane, respectively.[23] As a first approximation, we follow such simple interpretation to compute the results shown in **Table 1** which depicts the estimated values for the first PC resonance mode $\omega_o$ [THz], its frequency shift $\delta\omega_o$ [THz] as more molecules are loaded and the associated mass variation $\delta m$ per molecule written in [g/mol] units. One, two, four and eight molecules indexed by "CR" (coronene) and "BP" (biphenyl) attach on the graphene drum labeled on the table by "GR". However in such atomically thin resonators, the interaction between the mat and the dopant plays an important role on its response since the loading mass can interfere on the intrinsic tension of the sensor. In reality, the adsorbates can impart extra tension on the membrane promoting then an upward frequency shift as depicted in our results. This also raises an important point regarding the use of graphene as NEMS. Since graphene interaction with foreigner objects is an issue here, its functional detecting area appears as a prominent parameter that must be carefully adjusted in order to render optimal sensory performances. Such peculiar detecting features could only be observed when one considers a full atomistic description – instead of phenomenological continuous models - as the one followed here.

Discussing in more details the results displayed in Table 1, we can see that the first resonance mode for coronene doping gradually moves to higher frequency ranges as more molecules are included but it shifts back when 8 molecules are placed. From the simulations it is possible to notice that 8 coronene molecules cover a considerable area of the graphene membrane characterizing a limit that we named as "full coating". Therefore such amount of coronene flat objects restrain some of the natural vibrations of graphene. In other words, additional nodal lines are imparted on the doped graphene membrane due to the presence of relatively wide coronene molecules. Adopting our harmonical expectation for the frequency-mass variation proportion, an excellent agreement with the real molar mass of coronene molecules (300.35 g mol$^{-1}$) is obtained when the graphene host receives two dopants. We can say that we identified a situation where the resonator is capable to provide a precise mass response. In this case, a 60 x 60 Å$^2$ graphene membrane doped with two adsorbed coronene molecules comprises the optimum ratio between the functional area of the sensor and amount of dopants.

The example of single coronene doping also consists of a state where the sensor is still not able to provide a good measurement and we identify it as "diluted regime". Firstly, the graphene membrane is still rather relaxed due to the low quantity of molecules. As a result, the membrane can only capture most of the thermal vibrations of the molecule. Secondly, coronene molecules present a regular flat topology which causes a rather weak effect on the graphene under single doping. Two molecules could then induce the perfect signal on the sensor. In contrast to coronene dopants, biphenyl (BP) objects induce an abrupt shift in frequency already in single doping case until it saturates around 0.04 THz no mattering how many molecules are loaded. Such slight frequency fluctuations as the sensor is loaded with impurities indicate a weak interaction between the biphenyl and the graphene.

The erratic shape of a single biphenyl compound yields strong and long dynamical movements along the graphene drum as demonstrated in **Figure 1** but this is even more evident for the single doping. This can give the sensor device the "wrong impression" that a large number of molecules are bouncing on it. For this reason, the presence of a single biphenyl prompted a high mass variation on the system of 631.0 g mol$^{-1}$. As more biphenyl molecules are included on the sensor, the direct interaction between them starts to restrain their dynamical trajectories. Their repulsion exerts a significant constraint on their kinetic vibrations, allowing the graphene detector to perceive their mass more efficiently. The situation containing four adsorbed molecules retains the best molar mass estimation of 158 .0 g mol$^{-1}$ per molecule, in comparison to its real value of 154.21 g mol$^{-1}$. Once more, we detected the optimum (sensor area)/(amount of impurities) ratio which the device can provide ultimate responses. Eventually, a "full coating" state is also reached as more molecules bounce on the graphene mat and again the sensor responds inaccurately.

Finally, in **Figure 4** we present the results obtained for our third example system, the fullerene. Fullerene doping brings a situation where the sensor is not able to give any mass output. This is due to the fact that this system deviates from harmonicity. This can be seen on the upper panels of the figure showing the trajectory projections on the two first principal components for the pristine (black curve) graphene and doped with one (red curve), two (green curve), and four (blue curve) buckyballs. In particular for the second principal component solution, the characteristic sinusoidal shape of the projections is severely disturbed. Therefore a mass detection based on $\delta\omega_o \sim \delta m_{eff}$ proportionality is not effective since it is not possible to decompose the harmonic modes of the PC projections. The complete signal of the sensor in response to fullerene doping can still be mapped onto current spectral functions containing all the available modes which is shown on the lower panels.

So far, we could demonstrate that the size of detection area with respect to the type of dopants matters in the case of implementing graphene membranes as nanosensor systems. In contrast to estimations obtained by continuous models, the full atomistic description reveals that too small graphene mats can fail in sensing the mass of the molecules. Such size dilemma was also observed experimentally.[23] Nonetheless the abilities of such membranes should not be underestimated since they are not limited to mass detection. As long as a proper dynamic formulation is followed, it is possible to verify that these materials can infer other properties of the adsorbed molecules and their interaction with the detector. We based on the idea that the molecules can imprint their "ID" on the membrane as the time goes by. The signature of the molecule can be read by calculating the molecular center of mass (CoM) motion and its statistical correlation.

**Figures 5**. (upper panel) presents the evolution in time of deviations in the center of mass ($\delta$C$o$M) along $\hat{z}$-direction (perpendicular to the graphene plane) for each doping molecule. The lower panel shows their respective Fourier transformations which can be interpreted as the "sound" emitted by each particular dopant and how graphene is codifying them. The well-defined circular shape of fullerenes results in a rather smooth dependency of its C$o$M with time. It oscillates almost uniformly and with large amplitudes, indicating that buckyballs mostly bounce on top of graphene in a monotonous way. This limited bouncing movements lead to a quite narrow spectral response of its C$o$M deviation, with most of its signal standing below 4.0 THz. Coronene molecule presents a noisier behaviour on their C$o$M variation, however it fluctuates around small amplitudes during most of its dynamic

- 10 -

trajectory. Its spectral function is richer in comparison to the fullerene example with sharp signals popping up along the entire frequency range. Its highest responses are slightly "blue-shifted", being located inside the frequency window of 0.0-4.3 THz. Our last example, the biphenyl molecule, is significantly noisy, with dynamical fluctuations reaching high amplitudes. Its spectral response manifests relevant signals until 4.8 THz followed by a short suppression between 4.8-6.0 THz as well as for fullerene. Somehow fullerene and biphenyl cannot be "heard" by the graphene at this frequency range. Above 6.0 THz the signal of biphenyl is recovered.

For a better illustration of such imprints, we show on **Figure 6. a** planar view of the dynamic trajectories already presented in Figure 1 drawn by four coronene (left panel) and biphenyl (right panel) molecules. The dynamic on-fly path of the coronene molecules reveals that they stand closer to the membrane, in opposite biphenyl is mostly repelled from the membrane. Such information about the interaction with the detector and the shape of the molecules were successfully captured by the C*o*M curves. As discussed previously, biphenyl with its twisted benzene rings trigger intense kinetic movements on the graphene-mat. As a consequence, the molecules avert from the membrane and their CoM deviation undergoes strong dynamical fluctuations. Coronene expresses its flat form in terms of a noisy CoM but it vibrates within low amplitudes. Finally, the buckyball and its characteristic bouncing movement on the graphene surface is evident through its extended oscillatory behaviour of the CoM. In addition, the three molecules reveal their most intense CoM signals at clearly distinguishable frequencies: 3.2 THz (fullerene), 4.1 THz (coronene), and 1.2 THz (biphenyl). These results confirm that each molecule can imprint a sort of "identification" on the graphene-ear.

In summary, our work underlines the possibility of using the graphene membrane as a nanoscale detector. We developed a complete theoretical framework based on classical molecular dynamic assumptions and elaborated a statistical analysis of the time-dependent trajectories subsequently mapped onto spectral current curves. Our results point out that such classical approximations for the molecule/membrane pairs is enough to demonstrate the promising sensing abilities of the device which can resolve molecules with closely related properties. At first glance, one might say that the performance of our graphene-based device is inferior in comparison to specifically oriented mass spectrometers. Nevertheless, what our calculations really demonstrate is that graphene membranes are able to effectively measure the mass of adsorbed dopants for intermediate coverages. In other words, the interplay between the size of the membrane and the concentration of the analyte can play a crucial role on their performance. We also show that it is feasible to detect specific "fingerprints" left by the molecules on the graphene-mat and they can be inferred by IR or even Raman spectroscopy. Depending on the shape of the molecule, distinct signals can be captured by the graphene flake as a result of the dynamical movements generated by the dopants on the its surface. We expect that several technological branches, particularly in the NEMS industry, could benefit from our graphene based drum: a nanoscale "*ear*" that can hear the "*sounds*" produced by other molecules.

**Model.**

*Molecular dynamic model*.

Our graphene membrane doped with three different molecules (fullerenes, coronenes and biphenyl) was treated via molecular dynamic (MD) simulations using classical module FIRST embedded in CP2k package[29]. The system's parametrization was written in CHARMM[24] format style based on AMBER99[25] force field approximation using VEGAZZ[26] and home-made computational codes. The dynamical trajectories are described by a classical Lagrangian, (r,p) with isothermal constrain condition given by Nose-Hoover thermostat[30] applied to a system containing N particles,

$$L(r,p) = \sum_{i=1}^{N} \frac{m_i}{2} s^2 \dot{r}^2 - U(r) + \frac{Q}{2} \dot{s}^2 - \frac{L}{\beta} \ln(s) \qquad (2)$$

where $\beta = 1/k_BT$ ($k_B$ the Boltzmann constant and T the temperature) and {r,p} defines the momentum and position space phase. The first two terms of the Lagrangian represent the difference between kinetic energy and the potential energy [$U(r)$] of the real system with N particles of mass $m_i$. The adopted potential energy $U(r)$ follows a classical parametrization written in terms of two main contributions: bonded and non-bonded terms. The bonded elements involve 2-, 3-, and 4-body interactions of covalently bonded atoms which count for the harmonic, angular and dihedral angular motions of the atoms, respectively. The non-bonded term is simply modeled via long-ranged Lennard-Jones potential,

$$U_{NN}(r) = 4\varepsilon \left[ \left(\frac{\sigma}{r}\right)^6 + \left(\frac{\sigma}{r}\right)^{12} \right] \quad (3)$$

being $\epsilon$ the depth of the potential, $\sigma$ a finite distance which determines where the inter-particle potential is zero and r the interatomic distance. This formulation is based in a non-Hamiltonian treatment, hence, an additional set of phase space coordinates, {s, ṡ}, is introduced in the description associated with an effective mass Q which determines the coupling between the reservoir and the system of interests. A dynamical average observable, A, can be computed following NVT canonical ensemble where $<A(r,p/s)>_{Nose} = <A(r,p')>_{NVT}$ being the scale factor $L = 3N+1$ and $p' = p/s$. Principal component analysis (PCA) was subsequently adopted to project the obtained dynamical trajectories onto a new uncorrelated basis set of vectors. The obtained MD trajectories were fitted to one reference value for calculating the variance-covariance matrixes of the interatomic fluctuations. The matrixes were diagonalized, leading to a set of uncorrelated harmonic distribution functions, corresponding to the principal components. The principal component (PC) modes were finally arranged in descending order with respect to their amplitudes and up to four components were selected for analysis. PCA procedure was followed by using GROMACS[27] tools and the final data was visualized with VMD[28] package.

*Damped current ideas.*

The harmonic real space atomic translocations clearly visualized in **Figure 2(c)** led us to adopt a time-dependent continuous model to map the spectral current response of the pristine and doped graphene-mat. It is worth mentioning that we are not interested in quantitative values for the current itself but rather on the resonance frequency shifts caused by the presence of the molecules. Therefore, an effective transport model through a single channel suites perfectly our needs. Our first assumption is to assume that charges can ballistically flow through the graphene flake. The characteristic "wavy" shape of the membrane's vibrational modes clearly distinguishable after PCA treatment, allows us to consider that our system obeys the same "mathematics" of a driven RLC circuit. To simulate such harmonical charge flowing through the system, the coupling between the graphene and the electrodes ($\tau$) is considered to follow an harmonical dependency in time as $\tau \rightarrow \tau + \tau \sin(\omega t + \phi)$, being $\omega$ the frequency of the mode and $\phi$ its dephasing angle. The analytical expression for the electronic transmission at a certain energy of a single-level system [$\epsilon_0$] attached to two symmetric electrodes can be written as

$$T(\xi) = \frac{\tau^2}{(\xi - \varepsilon_o - \sigma)^2 + (\gamma/2)^2} \tag{4}$$

The metallic leads are represented by the self-energy term $\Sigma = \sigma - i\gamma/2$ where its real part, $\sigma$, can shift the level up or down and its imaginary part, $\gamma$, accounts for the level broadening. Assuming the harmonical time-dependency for $\tau$ and using Landauer-Bütticker equation, $I(\omega,t) = \int T(\xi',\omega,t)f(\xi')d\xi'$ (being $f(\xi')$ - the Fermi-Dirac distribution) we obtain

$$I = I_o + 2I_o \sin(\omega t + \varphi) + O[(\ldots)^2] \tag{5}$$

with $I_o = \int T(\xi')f(\xi')d\xi'$. **Expression 5** is also a solution for the second order differential equation which describes the current flowing along a RLC driven circuit. Its solution written in terms of frequency is given by where $p = (\omega - \omega_{ext})/\alpha$, $\alpha$ is the attenuation parameter of the circuit, the dephasing angle is related with the frequency shift as $\tan(\phi) \sim 1/p$, and $\{\phi, \omega_{ext}\}$ are the amplitude and the driven frequency, respectively, of an external pumping field expressed by $V(t) = E\cos(\omega_{ext} t)$. Within this assumption one can easily arrive to **equation 1**.

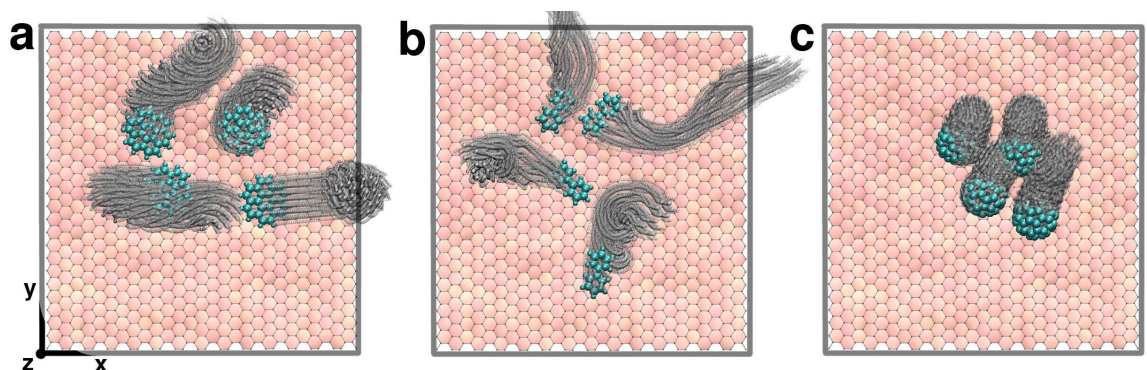

**Figure 1.** Dynamic paths developed by four (a) coronene, (b) biphenyl and (c) fullerene molecules added on the graphene plate. Each snapshot which delineates the sequence of shadows completes a representative 25 ps trajectory with 10 fs sampling between frames.

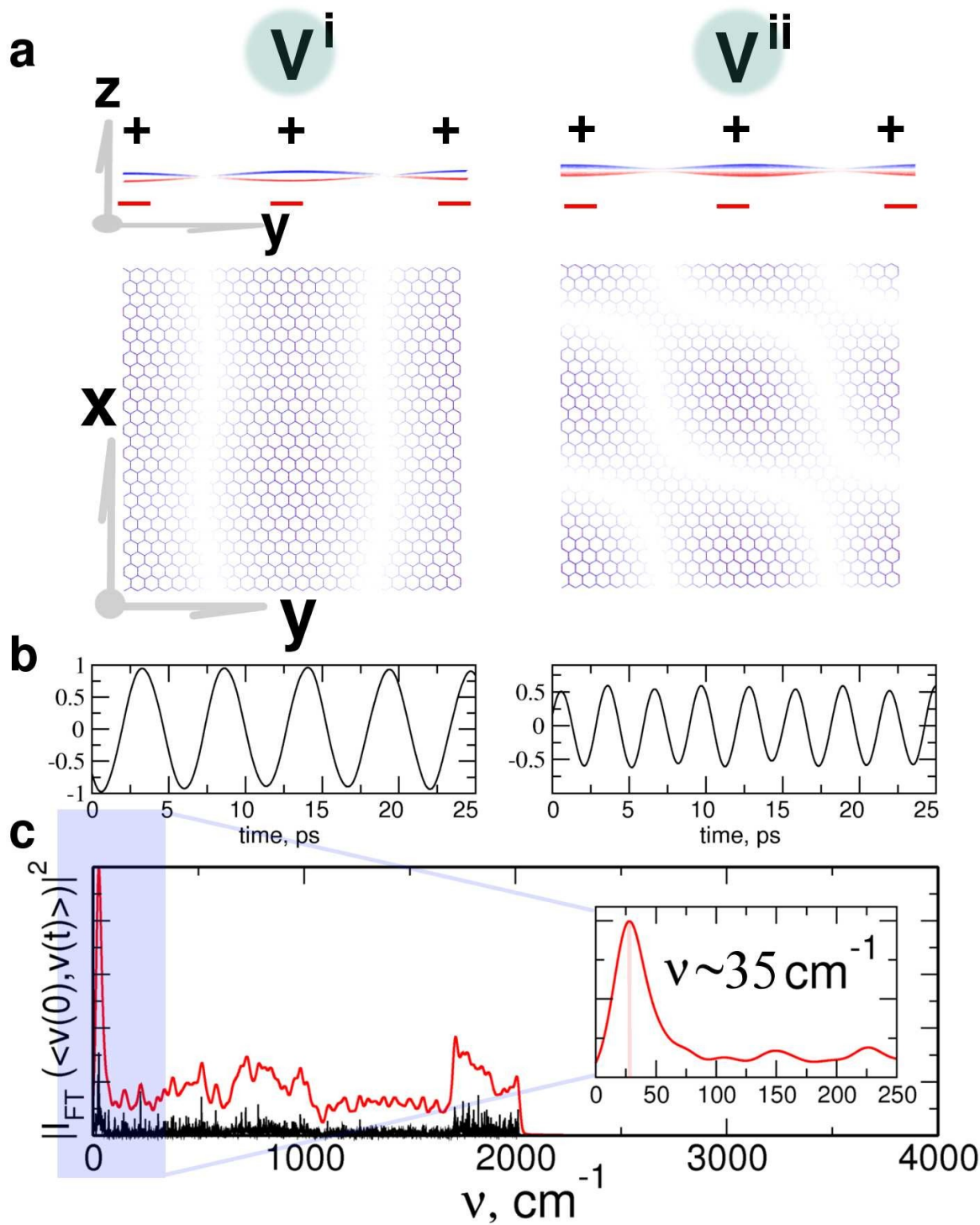

**Figure 2.** Molecular dynamic results obtained for a pristine graphene plate. (a) First and second vibrational modes of a graphene membrane. Discriminate positive and negative displacements of the atoms with respect to the graphene plane are represented by + and - signals. (b) Trajectory projection on the two first principle component vectors labeled by $V^i$ (left panel) and $V^{ii}$ (right panel). (c) Vibrational density of states taken from the velocity autocorrelation function. The inset blows up the range of lower frequencies. Characteristic positive and negative atomic displacements shown in panel (a) are distinguished by red and black curves.

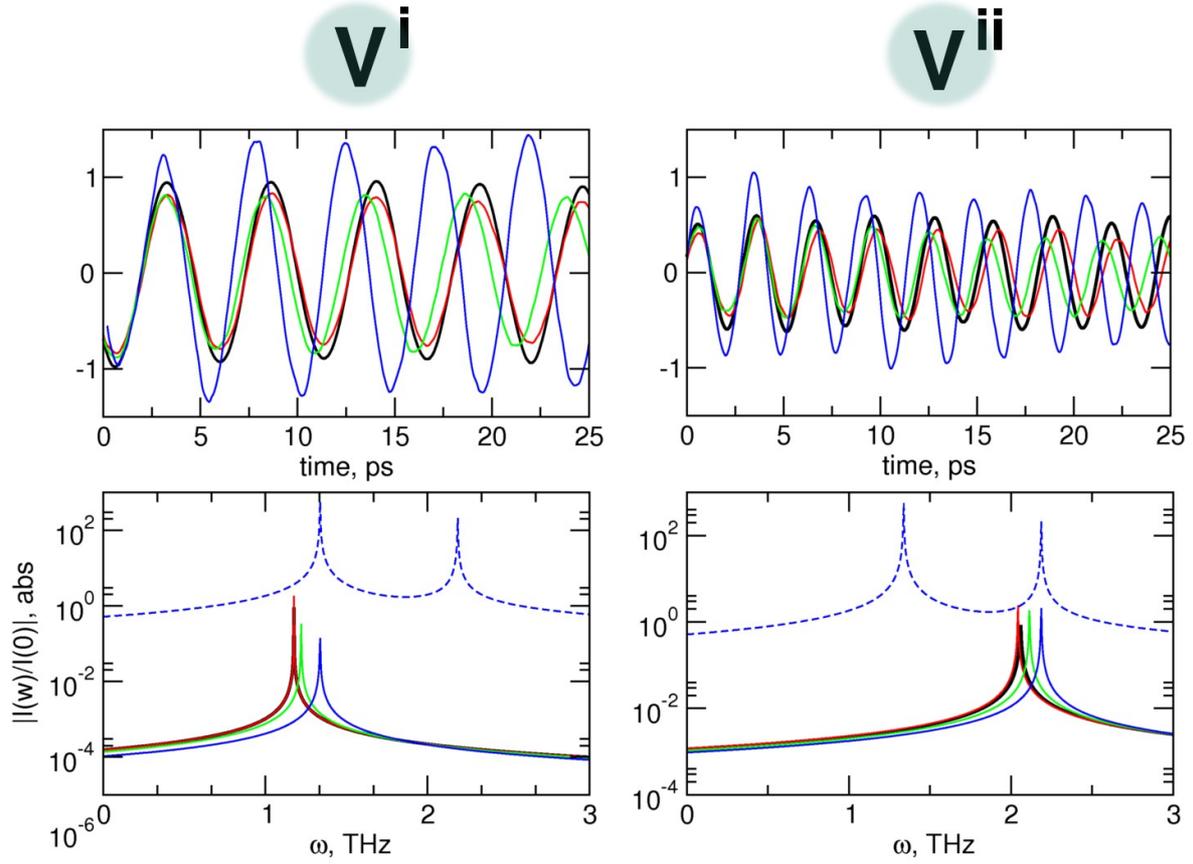

**Figure 3.** (upper panel) Trajectory projections on two first principal component vectors ($V^i$ and $V^{ii}$) for the graphene membrane on its pristine form (black) and doped with one (red), two (green), and four (blue) coronene molecules. (lower panel) Respective current curves [$I(\omega)/I(0)$] obtained from our harmonical model. The blue dashed line is the sum over the two components for the case where four molecules are added on the graphene.

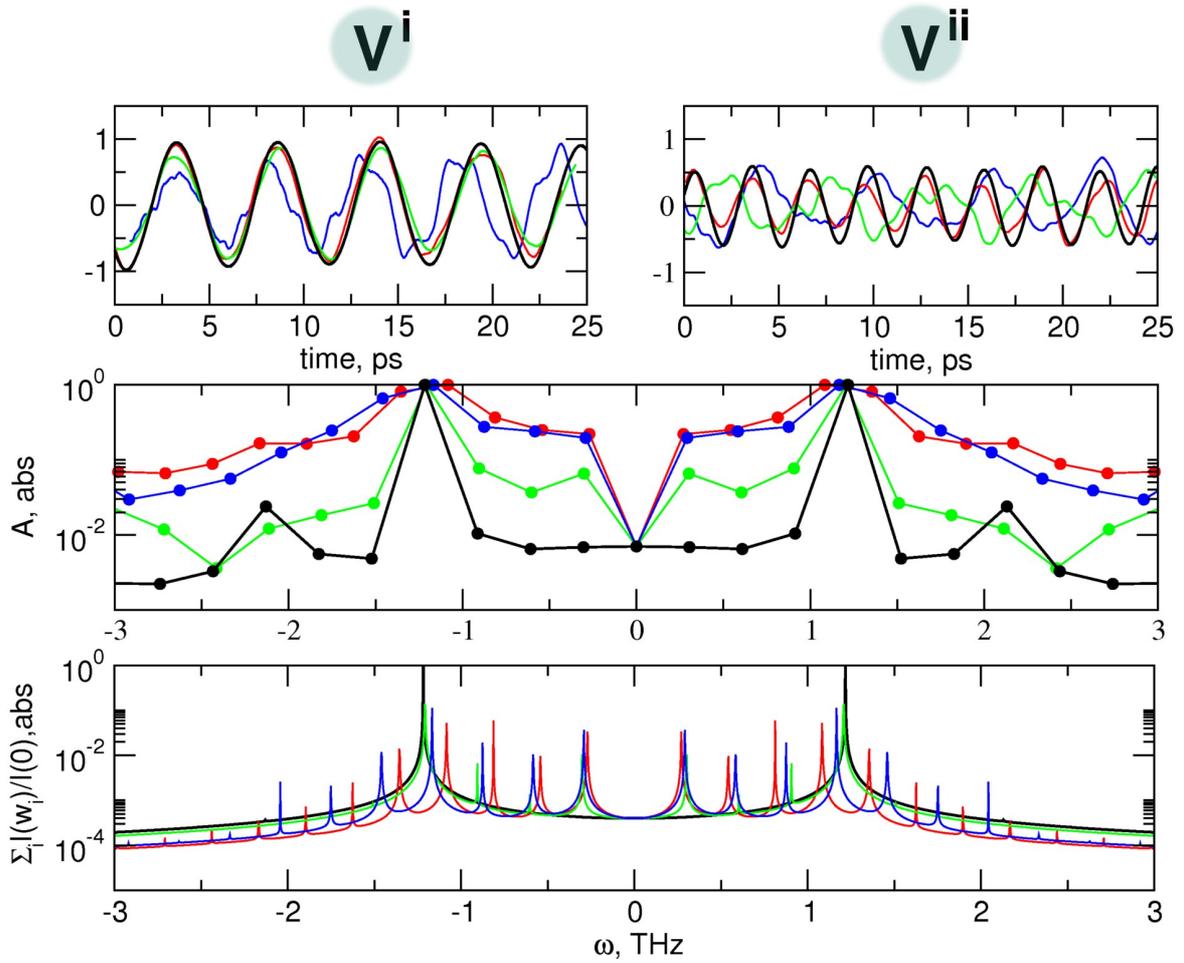

**Figure 4.** (Upper panel) Trajectory projections on two first principal component vectors ($V^i$ and $V^{ii}$) for pristine graphene layer (black) and doped with one (red), two (green), and four(blue) fullerene molecules. (Middle panel) Spectral function taken from the trajectory projections performed over the first principal component vector. (Lower panel) Current response behavior summed over all the harmonical modes extracted from equation 6.

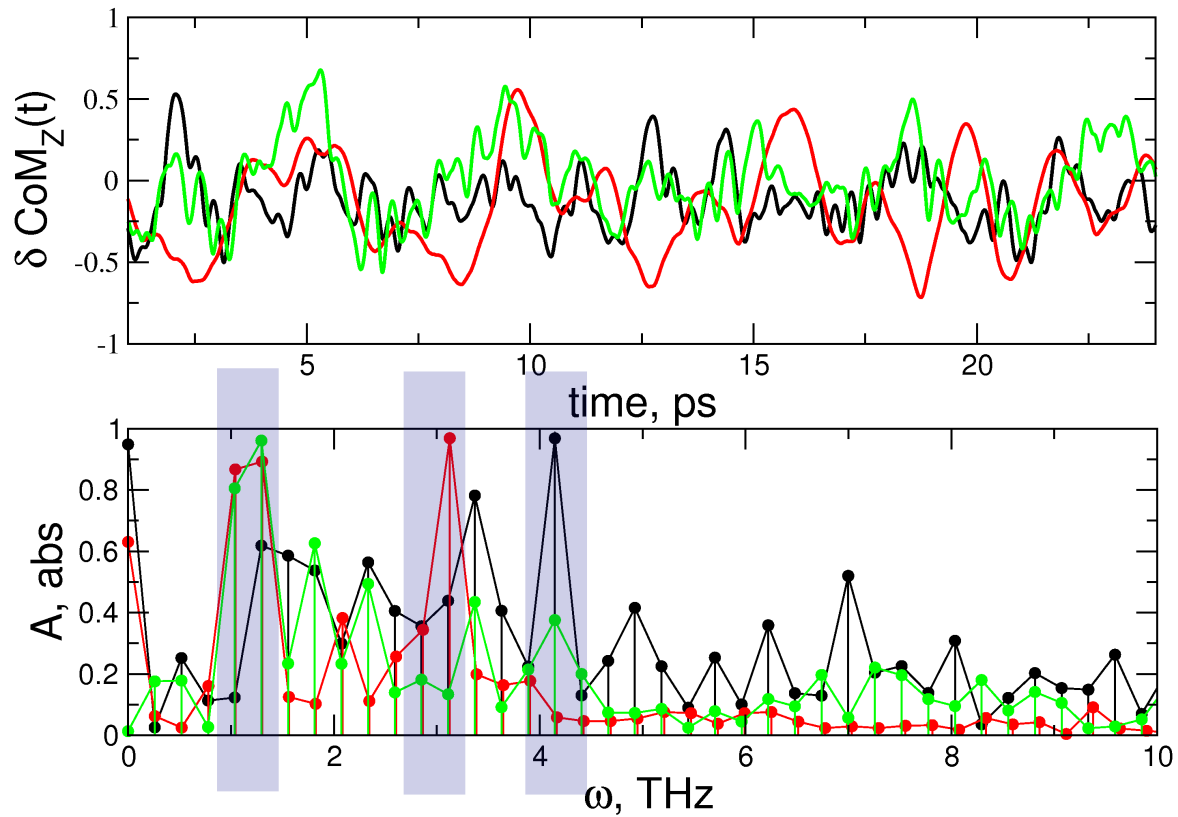

**Figure 5.** (upper panel) Center of mass variation ($\delta$ CoM) projected along $\hat{z}$-component which points perpendicularly to the graphene plane as a function of time. (lower panel) Normalized Fourier Transformation of the $\hat{z}$-component CoM deviation. Black, red and green curves correspond to the graphene membrane doped with coronene, fullerene, and biphenyl molecule.

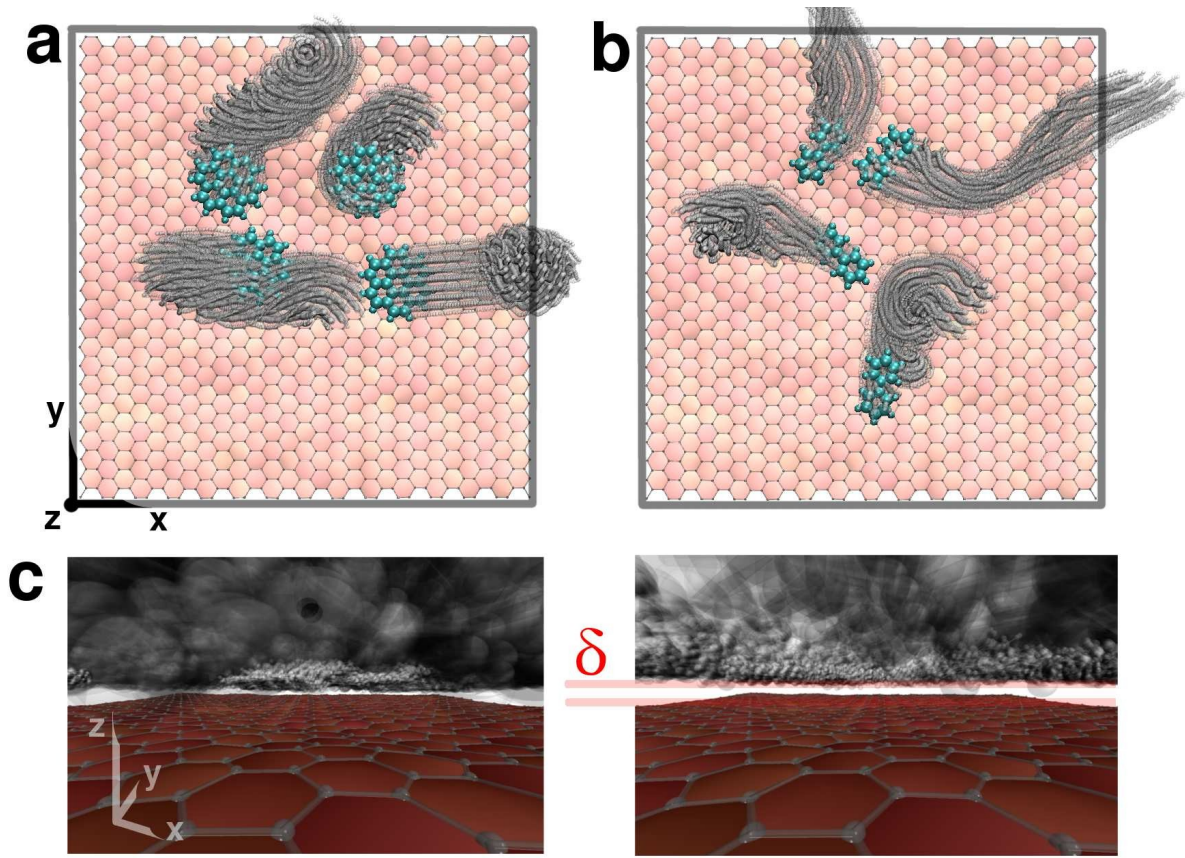

**Figure 6.** (Upper panels) Visualization of the dynamic paths developed by four (a) coronene and (b) biphenyl molecules deposited on the graphene membrane. (Lower panels) Planar view of the trajectory clouds drawn by the four (c) coronene and (d) biphenyl molecules on the membrane. $\delta$ evidences the more pronounced separation between biphenyl objects and the graphene.

**Table 1.** Resonance frequency ($\omega_0$), and variation ($\delta\omega_0$) and molecular mass variation $\delta m$ obtained for pristine graphene ("GR") and doped with coronene ("CR") or biphenyl ("BP") molecule

| Structure | GR | +1 CR | +2 CR | +4 CR | +8 CR |
|---|---|---|---|---|---|
| $\omega_0$ [THz] | 1.176 | 1.179 | 1.220 | 1.337 | 1.200 |
| $\delta\omega_0$ [THz] | 0.000 | 0.003 | 0.044 | 0.161 | [a] |
| $\delta m_{eff}$ [g/mol] | - | 43.0 | 314.0 | 575.0 | [a] |
| Structure | GR | +1 BP | +2 BP | +4 BP | +8 BP |
| $\omega_0$ [THz] | 1.176 | 1.221 | 1.210 | 1.221 | 1.18 |
| $\delta\omega_0$ [THz] | 0.000 | 0.044 | 0.040 | 0.044 | [a] |
| $\delta m_{eff}$ [g/mol] | – | 631.8 | 287.0 | 158 | [a] |

[a] Can't be mapped to harmonic picture

**What music do molecules play?** The molecular dynamics inside to the molecular level mechanical sensors.

S. M. Avdoshenko, C. G. Rocha, G. Cuniberti

Nanoscale ear drums: Graphene based nanoscale sensoric

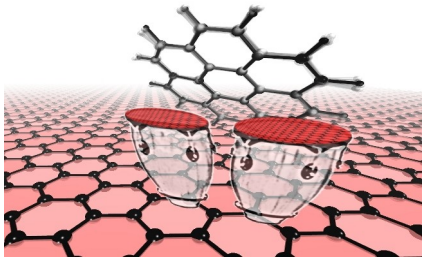